\documentclass[conference,10pt]{IEEEtran}
\IEEEoverridecommandlockouts
\usepackage{cite}
\usepackage{amsmath,amssymb,amsfonts}
\usepackage{algorithmic}
\usepackage{graphicx}
\usepackage{textcomp}
\usepackage{xcolor}
\usepackage{subcaption}
\usepackage{url}
\usepackage{booktabs}
\usepackage{balance}

\usepackage[absolute,showboxes]{textpos}

\usepackage{tikz}
\usetikzlibrary{arrows,arrows.meta,external,colorbrewer,decorations.markings,external,patterns,shapes}
\tikzexternaldisable

\usepackage{pgfplots}
\usepgfplotslibrary{groupplots,colormaps,colorbrewer,statistics}
\usepackage{pgfplotstable}
\pgfplotsset{compat=1.14}

\usepackage{pifont}
\newcommand{\cmark}{\ding{51}}%
\newcommand{\xmark}{\ding{56}}%

\usepackage{xspace}
\newcommand{\eg}{\textit{e.g.,}~}

\newcommand{\cf}{\textit{cf,}~}
\newcommand{\etc}{\textit{etc.}~}
\newcommand{\one}{({\em i})\xspace}
\newcommand{\two}{({\em ii})\xspace}
\newcommand{\three}{({\em iii})\xspace}
\newcommand{\four}{({\em iv})\xspace}

\makeatletter
\renewcommand{\paragraph}[1]{\vspace*{0.03in}\noindent{\bf #1.}\hspace{0.25ex \@plus1ex \@minus.2ex}}
\newcommand{\paragraphb}[1]{\vspace*{0.03in}\noindent{\bf #1}\hspace{0.25ex \@plus1ex \@minus.2ex}}
\makeatother

\begin{document}

\title{IoT Content Object Security with OSCORE and NDN: A First Experimental Comparison}

\author{\IEEEauthorblockN{Cenk G\"undo\u{g}an}
\IEEEauthorblockA{\textit{HAW Hamburg}\\
\small cenk.guendogan@haw-hamburg.de}
\and
\IEEEauthorblockN{Christian Ams\"uss}
\IEEEauthorblockA{\textit{~} \\
\small christian@amsuess.com}
\and
\IEEEauthorblockN{Thomas C. Schmidt}
\IEEEauthorblockA{\textit{HAW Hamburg} \\
\small t.schmidt@haw-hamburg.de}
\and
\IEEEauthorblockN{Matthias W\"ahlisch}
\IEEEauthorblockA{\textit{Freie Universit\"at Berlin} \\
\small m.waehlisch@fu-berlin.de}
}

\maketitle

\setlength{\TPHorizModule}{\paperwidth}
\setlength{\TPVertModule}{\paperheight}
\TPMargin{5pt}
\begin{textblock}{0.8}(0.1,0.02)
     \noindent
     \footnotesize
     If you cite this paper, please use the Networking reference:
     C. G{\"u}ndo\u{g}an, C. Ams\"uss, T.~C. Schmidt, M. W\"ahlisch. IoT Content Object Security with OSCORE and NDN: A First Experimental Comparison. In \emph{Proc. of 19th IFIP Networking}, IEEE Press, 2020.
\end{textblock}

\begin{abstract}
  The emerging Internet of Things (IoT) challenges the end-to-end transport of the Internet by low power lossy links and  gateways that perform protocol translations. Protocols such as CoAP or  MQTT-SN are degraded by the overhead of DTLS sessions, which in common deployment protect content transfer only up to the gateway.  	
  To preserve content security end-to-end via gateways and proxies, the IETF recently developed Object Security for Constrained RESTful Environments (OSCORE), which extends CoAP with content object security features commonly known from Information Centric Networks (ICN).

  This paper presents a comparative analysis of protocol stacks that protect request-response transactions.
  We measure protocol performances of CoAP over DTLS, OSCORE, and the information-centric Named Data Networking (NDN) protocol on a large-scale IoT testbed in single- and multi-hop scenarios.
  Our findings indicate that (a) OSCORE improves on CoAP over DTLS in error-prone wireless regimes due to omitting the  overhead of maintaining security sessions at endpoints, and (b) NDN attains superior robustness and reliability due to its intrinsic network caches and hop-wise retransmissions.
\end{abstract}


\section{Introduction}\label{sec:intro}

The Internet  design follows an end-to-end principle~\cite{src-eeasd-84}, which strongly shaped  its transport layer. Transport sessions shall establish directly  between applications without intermediaries. Secure networking is dominantly deployed by (datagram) transport layer security  (D)TLS. (D)TLS interception, however, breaks the end-to-end paradigm from a security perspective.  At the same time, a growing number of use cases demands for application layer gateways and transport assistance, which both hinder end-to-end session security. 

The Internet of Things (IoT) emerges with massive deployments of constrained devices that are shielded behind application gateways.  These gateways translate between  the Constrained Application Protocol (CoAP) over DTLS and HTTPS, or the Message Queuing Telemetry Transport for Sensor Networks MQTT-SN over DTLS and MQTT over TLS, which require re-authentication and re-encryption. In addition, cryptographic overhead burdens the constrained nodes in its low end wireless transmission systems and makes it hard to maintain security sessions for all the small data transfers. 
The IoT is thus a use case against end-to-end session security.

Adding security credentials to content objects instead of transmission channels is an orthogonal approach to secure communication on the Internet. It changes the session-centric paradigm by adding authentication and encryption (if desired) to each data chunk, which in turn allows for content caching and transport translation at gateways, while preserving data security properties. Information Centric Networking first introduced content object security on the network layer for the sake of ubiquitous caching. Recently, the IETF Core working group released OSCORE, which extends the IoT ecosystem to content object security. 

In this paper, we present and comparatively evaluate the full solution space for secure content transmission in the IoT. Starting from a problem statement and related protocol work in Section~\ref{sec:background}, we present a comprehensive set of implementations within the RIOT~\cite{bhgws-rotoi-13} networking subsystem in Section~\ref{sec:implementation}. Theoretical evaluations follow in Section~\ref{sec:evaluation-theory}. Our network experimentation on a large-scale testbed are discussed in Section~\ref{sec:eval} along with various results that indicate significant performance improvements over CoAP/DTLS by OSCORE as well as NDN. Section~\ref{sec:conclusion} concludes with an outlook.

\section{The Problem of Securing IoT Content and Related Protocol Work}\label{sec:background}

\begin{figure*}
  \centering
  \includegraphics{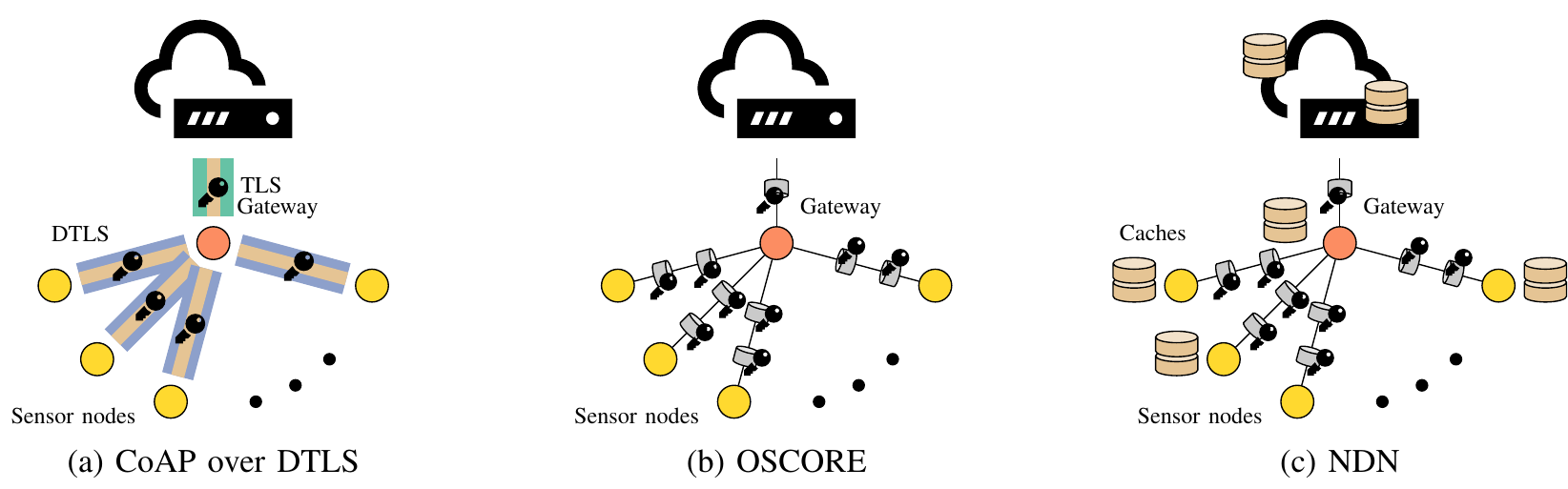}
  \caption{Typical deployment setups for CoAP over DTLS, OSCORE, and NDN in the IoT.}%
  \label{fig:overview}
\end{figure*}

\subsection{Problem Statement}

The Internet of Things is evolving to connect numerous, often constrained devices that regularly exchange massive amounts of data. Authenticity and possibly confidentiality of information is of vital interest in a wide range of applications. The problem, though, is that low-end devices need to optimize resources and thus need to minimize cryptographic operations and state while (re-)transmitting packets.

At the same time, low-power lossy networks frequently experience packet loss and require retransmissions---multihop transfers often significantly challenge  these error-prone regimes~\cite{gklp-ncmcm-18}.  Overhead from cryptographic credentials or signaling security sessions consumes additional energy and may quickly become critical for these low-power devices. 

On the contrary, trust relationships in the IoT may be heterogeneous and change with varying deployment settings. While the exchanging endpoints are often widely distributed (\eg sensors and a cloud), IoT gateways often need to translate between protocols. If translators are required to re-authenticate and re-encrypt, all communicating parties must pre-establish trust with the IoT gateways in place. This will be a major problem in provider-bound deployments such as 5G.

\subsection{Transport Layer Security in the IoT}
Datagram Transport Layer Security (DTLS)~\cite{RFC-6347} closely follows TLS~\cite{RFC-8446} in terms of protocol behavior and  security guarantees.
Unlike its stream-oriented relative, DTLS adds facilities to operate in unreliable datagram environments.
It contributes a modified record layer that tolerates packet loss and message reordering.
To break up inter-record dependencies, DTLS bans stream ciphers and uses explicit sequence numbers in datagrams.
A cryptographic context thus spans exactly one record.
UDP is the prevailing transport in IoT deployments.
Compared to TCP, it exhibits no substantial protocol overhead and allows for implementations with low memory footprints.
Utilizing DTLS to secure existing application protocols such as CoAP and MQTT-SN hence appears to be the best logical choice---at least at first glance.

Concerns arose in recent studies that question the applicability of DTLS in large-scale IoT systems.
First, certain cryptographic challenges during handshake processes are infeasible.
While processing time for cryptographic operations diminish with hardware crypto modules, message sizes are  inflated.
Asymmetric key ciphers require handshake overhead and large payload sizes, which immensely boost handshake completion times to the order of seconds and minutes in multi-hop deployments due to packet fragmentation~\cite{kpk-cdcdr-16}.

The stateful session characteristic further comes at the cost of multicast capability,
since security contexts are identified by the classic 5-tuple  between two endpoints. 
Particularly in scenarios that involve device mobility and multi-homing, a generally accepted effort applies connection identifiers to  security channels---independent of the 5-tuple~\cite{draft-ietf-tls-dtls-connection-id}.
Figure~\ref{fig:overview} (a) illustrates a realistic deployment setup for CoAP over DTLS\@:
End-to-end security commonly terminates at the gateway to allow for protocol conversions, \eg to HTTPS over TCP\@.

\subsection{Content Object Security in the IoT}

OSCORE~\cite{RFC-8613} is a protocol extension to CoAP and addresses the terminating security issue at gateways.
Instead of securing sessions between endpoints, OSCORE protects entire CoAP messages and provides integrity, authenticity, and confidentiality on an object level.
The original CoAP message is thereby encapsulated as an authenticated and encrypted COSE~\cite{RFC-8152} object by an outer CoAP option.
In addition to cryptographic efforts, the protocol further includes countermeasures to prevent response delay and mismatch attacks.
A strong message binding between requests and corresponding responses is constructed with the use of identical identifiers in their authenticated components, which persist over retransmissions.
Replay windows allow for rearranged messages to be processed independently. Applications built on it use CoAP mechanisms like If-Match or the Echo~\cite{draft-ietf-core-echo-request-tag} option to protect against any ill-effects of rearranged messages.

OSCORE utilizes the request-response semantics of its underlying CoAP layer and an elaborate nonce construction to obtain compact response messages.
When combined with CoAP observation (continuous responses to a single request), OSCORE protects the sequence of notifications using its own sequence numbers.
When combined with CoAP block-wise transfer, it fragments large resources into pieces small enough for the end points to process in a single cryptographic operation without hindering further block-wise processing by proxies.
Unlike DTLS, OSCORE does not come with a built-in key exchange protocol, and relies on pre-shared keys.
A lightweight authenticated key exchange (LAKE~\cite{draft-ietf-lake-reqs}) is being developed as a companion protocol.

A major improvement over the conventional transport layer security concept is the ability to secure multicast messages.
CoAP supports a one-to-many group communication~\cite{RFC-7390} when used with UDP\@.
While DTLS fails to perform in multicast environments, the object security characteristic of OSCORE allows for protected requests and responses in these deployments~\cite{draft-ietf-core-oscore-groupcomm}.

Figure~\ref{fig:overview} (b) illustrates the envisioned deployment option.
Messages are cryptographically secured and despite protocol conversions on gateways, their properties stay intact while traversing up to cloud services.

\subsection{Content Security in the Information-Centric IoT}

Information-centric Networking~\cite{adiko-sind-12,xvsft-sinr-14}---a clean-slate approach of the  Future Internet initiatives---abandons the host-centric Internet paradigm in the favor of autonomous content, which allow for an unhindered  replication of authenticated data objects.
A decade of research has created a variety of ICN flavors that have three principles in common~\cite{gsksr-icnsf-11}: \emph{Decoupling of named content from hosts}, \emph{universal caching}, and \emph{content object security}.

Named Data Networking (NDN)~\cite{jstp-nnc-09} enjoys significant popularity and has been identified early as a candidate for low-end IoT edge networking~\cite{bmhsw-icnie-14}. An adaptation layer to the low power lossy wireless exists with ICNLoWPAN~\cite{gksw-innlp-19}. 
In contrast to the end-to-end stateless packet processing on the Internet, NDN utilizes a stateful, hop-by-hop forwarding fabric that decouples content objects from their locations and enables seamless on-path caching.
NDN follows a simple request-response scheme on the network layer using the two message types \emph{Interest} and \emph{Data}, each treated individually by the forwarding state machine. Figure~\ref{fig:overview} (c) illustrates the additional content caches that support content replication and local recovery from losses. 

NDN supports integrity and authenticity  as  protocol features by appending cryptographic  signatures to data packets.
While originally the intrinsic security only applied to data packets, the upcoming NDN protocol version allows for a signature inclusion in Interests.
Confidentiality is not supported on the protocol level, but left to the application to  encrypt content.


\section{Composable Network Stacks for Object Security in Challenged IoT Deployments}\label{sec:implementation}

The decision for a software platform that can cope with constrained IoT is crucial.
As we aim for maintainability and sustainability, we extend existing code bases instead designing and implementing from scratch.
As such, we utilize the open source IoT operating system RIOT~\cite{bghkl-rosos-18} and leverage the existing network stack architecture.
In course of our evaluations, we contribute and upstream improvements to the RIOT integrations of DTLS, OSCORE, and NDN.

\subsection{The RIOT Networking Subsystem}
\begin{figure}
  \centering
  \includegraphics{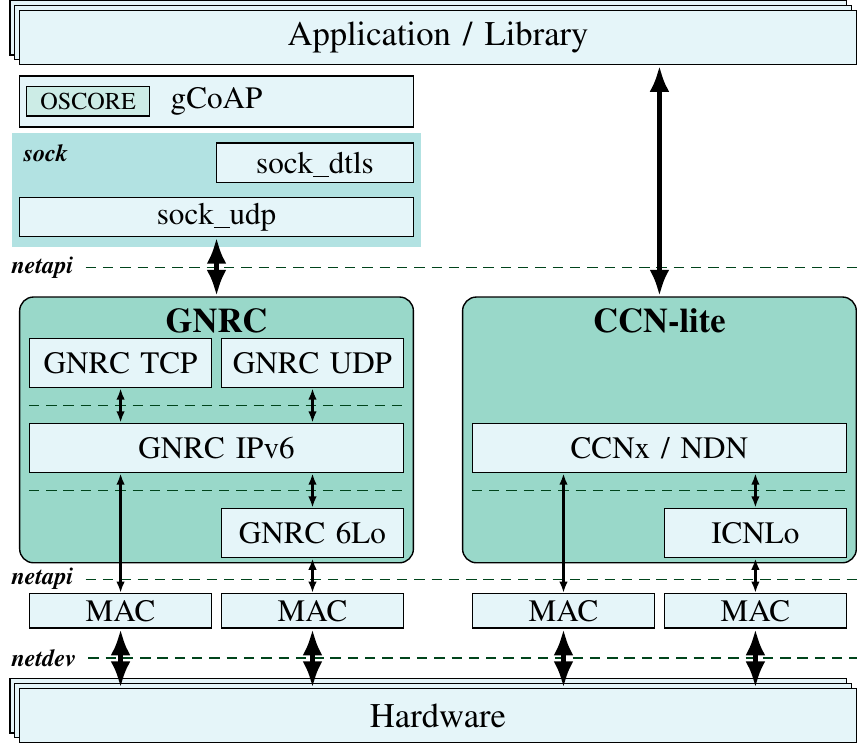}
  \caption{The RIOT networking subsystem.}%
  \label{fig:nwstack}
\end{figure}

The RIOT networking subsystem displays two interfaces to its externals (see Figure~\ref{fig:nwstack}): The application programming interface \texttt{sock} and the device driver API \texttt{netdev}.
Internal to stacks, protocol layers interact via the unified interface \texttt{netapi}, thereby defining a recursive layering of a single concept that enables interaction between various building blocks: 6lo with MAC, IP with routing protocols, transport layers with application protocols, \etc
This grants enhanced flexibility for network devices that come with stacks integrated at different levels.

\subsection{CoAP Over DTLS}
gCoAP is the feature-rich native CoAP implementation of RIOT\@.
It implements the server-side and client-side, it supports the most common methods GET, POST, PUT, DELETE, it handles confirmable messages, and it allows for observing resources.
As depicted in Figure~\ref{fig:nwstack}, gCoAP uses the sock API\@.
On the north-bound it attaches to \texttt{sock\_udp} and \texttt{sock\_dtls}, which makes it completely network stack agnostic.
In the default configuration, the native 6LoWPAN network stack of RIOT, GNRC, provides the south-bound implementation of \texttt{sock\_udp}.
The DTLS counterpart is provided by the external package tinyDTLS\@.
It follows a threadless design and depends on events, which are handled by the sock layer within the gCoAP thread.
This DTLS setup supports two ciphersuits: \one pre-shared authentication and key exchange using AES encryption in Counter with CBC-MAC (CCM)~\cite{RFC-3610} mode with a 128-bit key length, and \two an Elliptic Curve Cryptography (ECC) based AES-CCM with an Elliptic Curve Diffie-Hellman Ephemeral (ECDHE) key exchange.

\subsection{CoAP With OSCORE}
We provide \texttt{libOSCORE}\footnote{\url{https://gitlab.com/oscore/liboscore}} as an implementation of the OSCORE~\cite{RFC-8613} model that integrates into RIOT\@.
Unlike other approaches, \eg  a yet to be mainstreamed contiki implementation\footnote{\url{https://github.com/Gunzter/contiki-ng/tree/oscore_12}} and c\_OSCORE\footnote{\url{https://github.com/Fraunhofer-AISEC/c_OSCORE}} on top of Zephyr, \texttt{libOSCORE} focuses on portability across different CoAP libraries and provides replay protection.

Distinct features of \texttt{libOSCORE} are its handling of the request-response correlation data and its zero-copy approach.
In the former, Partial IVs are consistently passed by reference. They carry a flag indicating first use, which gets invalidated by consumers.
This allows leveraging OSCORE optimizations for safe REST operations.
In memory management, \texttt{libOSCORE} expects its user to provide suitable memory locations and provides struct definitions to make that portable.
This saves execution time, RAM and ROM at the cost of some implementation complexity on the user side.
It allows processing of messages from the receive-buffer in a single reading pass after in-place decryption and without dynamic memory allocation.

In CoAP libraries that build and read their messages in buffers (\cf gCoAP),
integration of \texttt{libOSCORE} happens in two stages: \one basic integration, and  \two full integration.

The basic integration describes the most elementary way of interacting with \texttt{libOSCORE}\@.
It only requires a mapping of certain CoAP operations and cryptographic primitives.
Applications that use this mode directly access OSCORE objects and steer every step of the encryption and decryption process for each packet.
Generally, usage of this mode is tedious and error-prone and therefore discouraged for user applications.
On the other hand, a basic integration allows for a full control of OSCORE internals, which can be leveraged to perform protocol optimizations by libraries or protocol extensions.

The full integration requires a functional basic integration as prerequisite.
At that stage, \texttt{libOSCORE} messages are used as backends for the native CoAP library.
Application code is identical for the unprotected and OSCORE-protected case, and thus tightly coupled to the native CoAP implementation.
The CoAP library dispatches operations on messages through \texttt{libOSCORE} atop of, or directly through the transport protocol, depending on the application's configuration (\eg the choice of a security context for a message) or presence of the OSCORE option.
This mode of operation is the recommended way of building user applications, as APIs hide security operations and prevent security breaches due to a misuse of OSCORE internals.
The stack in Figure~\ref{fig:nwstack} shows the full integration state where applications interact only with gCoAP\@.

An intermediate integration is available in \texttt{libOSCORE} for cases when full integration is unfeasible with a particular library or simply incomplete.
At that level, an additional library provides code to orchestrate and simplify cryptographic procedures.
This mode is most suited for narrow-purpose helper libraries up to full-fledged REST frameworks, which generally provide their own APIs towards user applications.

\subsection{Named Data Networking}
CCN-lite~\cite{ccn-lite} is a lightweight NDN forwarder, which supports all primary features:
in-network caches, hop-wise retransmissions, request aggregation along paths, and multi-source, multi-destination forwarding.
It runs on a variety of hardware platforms---ranging from commodity hardware to embedded devices.
While the core forwarder is self-contained and platform independent, adaptors provide access to the system communication API\@.
CCN-lite is integrated into RIOT as an external package.
It contributes a RIOT adaptor, which hosts its own thread and translates between CCN-lite messages and netapi packets.


\section{Theoretical Evaluation}\label{sec:evaluation-theory}

Performance measures such as security properties largely differ for each protocol configuration.
In the following, our performance assessment considers protocol design choices and thus provides insights that are independent of specific deployments.
We focus on four protocol compositions:
\one CoAP (Protected) with encrypted and authenticated response payload as baseline implementation.
\two CoAP over a secured DTLS~1.2 session.
\three OSCORE to provide object security for request and response messages.
\four NDN (Protected) using signed Data messages and encrypted as well as authenticated content.

\subsection{Security Properties}

\paragraphb{CoAP (Protected)} exhibits the weakest security properties in our comparison:
While it uses an authenticated encryption for the \emph{payload}, it does not provide any security measures for the actual CoAP messages to protect CoAP signaling.
Protocol headers are prone to tampering and messages are susceptible to interception as well as packet delay attacks.
These shortcomings make the binding of requests to correct responses fragile.
The inability to map responses to particular requests is especially dangerous in cases when resources publish mutable content~\cite{draft-ietf-core-echo-request-tag,draft-mattsson-core-coap-actuators}.
Consequently, even in the case when the payload is secured, delayed and replayed messages can affect the state machine on the client and server.
Since the message headers are not protected against confidentiality attacks, this configuration easily leads to privacy concerns.
Plaintext requests will contain resource URIs, which typically help to identify sensitive application information and therefore potentially leak private data.
Responses may not include resource URIs, but included tokens unambiguously identify potentially intercepted requests and thus their resource URIs.

\paragraphb{CoAP over DTLS} is the common method for securing message transmissions in an IoT network.
DTLS provides integrity, authenticity, and confidentiality for UDP datagrams within sessions based on pre-established private keys.
It operates below the application layer and inherently takes CoAP requests as well as responses into consideration.
A drawback from this layering, however, is that the DTLS record layer is not aware of CoAP semantics.
This introduces a twofold problem:
First, this configuration suffers from the same request-response binding issues when messages are delayed and replayed~\cite{draft-mattsson-core-coap-actuators} while recent mitigations~\cite{draft-ietf-core-echo-request-tag} are not deployed yet.
Second, end-to-end security terminates at gateways in usual IoT setups when protocol conversions from CoAP to HTTP take place.
Minimal DTLS implementations commonly provide the lightweight DTLS cipher suite TLS\_PSK\_WITH\_AES\_128\_CCM\_8~\cite{RFC-6655}, which does not provide perfect forward secrecy.
Adaptations~\cite{RFC-8442}, allow for the combination of existing cipher suites with the Ephemeral Elliptic Curve Diffie-Hellman key agreement protocol.

\paragraphb{OSCORE} achieves a secured communication by protecting request and response messages on CoAP level.
This is in contrast to CoAP over DTLS that establishes secure channels between endpoints.
OSCORE provides integrity, authenticity, and confidentiality by nesting the actual CoAP message as an authenticated and encrypted payload, interleaving information relevant to routing and retransmission in the unprotected outer parts.
This layer hides sensitive information, such as the resource path and the CoAP method of the original message.
Furthermore, the security of inner messages stays intact across protocol translations on gateways (\eg from CoAP to HTTP/S).
OSCORE provides a strong request-response binding with mechanisms like sequence counters and sliding windows, which renders many attacks ineffective.
The original specification is missing a key exchange protocol and thus does not provide perfect forward secrecy.
Adaptations~\cite{draft-selander-lake-edhoc} allow for an Ephemeral Diffie-Hellman over COSE\@.

\paragraphb{NDN} authenticates response messages between arbitrary endpoints without the need for session state.
While application payload can be encrypted, NDN does not provide confidentiality for message headers.
Moreover, NDN reduces security features to response messages only\footnote{Specification v0.3 is in progress and adds security features to Interests}.
Names are an integral part of the NDN forwarding fabric and may contain sensitive application information.
Thus, privacy concerns arise from plaintext names in NDN messages.
An encryption or obfuscation of names inevitably affects the routing system and adds an exhaustive overhead.
Unlike the CoAP variants, NDN follows the principle of immutable content: A specific name invariably points to the same content object.
This property reduces the attack surface and desensitizes applications to delayed and replayed messages.

We summarize the observed advantages and drawbacks of the discussed protocol schemes in Table~\ref{tab:secprops}, with a strong focus on the actual protocol behavior rather than on  application payload security.

\begin{table}
  \centering
  \caption{Summary of security properties for each protocol configuration.
    (\cmark) indicates optional specifications, which are unavailable in the used implementations.}%
  \label{tab:secprops}
  \begin{tabular}{@{}lccccc@{}} \toprule
    & \multicolumn{3}{c}{\textbf{CoAP}} & \textbf{NDN} \\
    \cmidrule(lr){2-4}
    & Protected & DTLS & OSCORE & Protected \\ \midrule
    \textbf{Request Message}\\
    Integrity & \xmark & \cmark & \cmark & (\cmark)\\
    Authenticity & \xmark & \cmark & \cmark & (\cmark)\\
    Confidentiality & \xmark & \cmark & \cmark & \xmark\\
    \textbf{Response Message}\\
    Integrity & \xmark & \cmark & \cmark & \cmark\\
    Authenticity & \xmark & \cmark & \cmark & \cmark\\
    Confidentiality & \xmark & \cmark & \cmark & \xmark\\
    \textbf{Attack Resiliency}\\
    Replay Insensitivity & \xmark & (\cmark) & \cmark & \cmark\\
    Perfect Forward Secrecy & \xmark & (\cmark) & \xmark & \xmark\\
    \bottomrule
  \end{tabular}
\end{table}

\subsection{Security Message Overhead}\label{subsec:secoverhead}

In all protocol configurations, security extensions add message overhead and consequently affect transmission times.
Notably for IEEE~802.15.4, inflated messages easily increase media access times by a few milliseconds, whereas computational overhead in common IoT network stacks is in the range of microseconds~\cite{lkhpg-cwemr-18}.
We now quantify the overhead in terms of packet size which is introduced by the different security extensions.
In Section~\ref{sec:packet-structures}, we will put this into perspective with respect to the common CoAP and NDN packets.

\begin{figure*}
  \centering
  \includegraphics{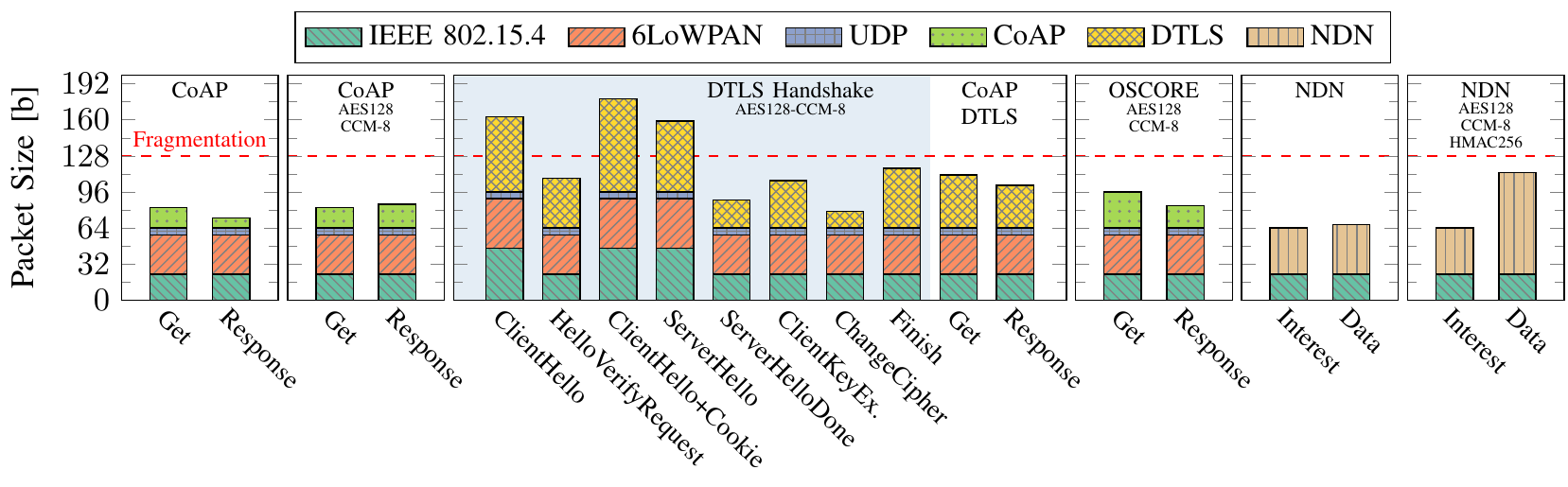}%
  \caption{Packet structures of control- and data-plane packets for each protocol configuration.}%
  \label{fig:packetstructure}
\end{figure*}

CoAP (Protected) and NDN (Protected) do not add any message overhead to requests.
All configurations other than CoAP (Protected) add a structural overhead related to security.
DTLS includes 11 bytes for the DTLS~1.2 record layer in all datagrams, excluding the epoch field.
The NDN packet format uses flexible Type-Length-Value (TLV) fields to encode message headers.
Security related TLVs similarly account for 11 bytes overhead.
OSCORE exploits implicit information that results from a strong request-response binding and further utilizes a minimal CBOR representation.
This nets to a structural overhead of four and three bytes.

The security context identifier consists of two bytes for CoAP (Protected) and CoAP over DTLS\@.
In the former scenario, contexts are identified by the 2-byte key identifier within our payload, while the latter scenario uses a 2-byte epoch field in the record layer to denote a secured session.
OSCORE and NDN are able to reduce the length of small context identifiers.
OSCORE omits the security context in response messages and requesting devices must deduce it from the request state.

\begin{table}
  \centering
  \caption{Message overhead of security measures in bytes. Overhead does not apply to CoAP and NDN requests.}%
  \label{tab:overheads}
  \begin{tabular}{@{}lc@{\hspace{0.15cm}}cc@{\hspace{0.15cm}}cc@{\hspace{0.15cm}}cc@{\hspace{0.15cm}}cc@{}} \toprule
    & \multicolumn{6}{c}{\textbf{CoAP}} & \multicolumn{2}{c}{\textbf{NDN}} \\
    \cmidrule(lr){2-7}
    & \multicolumn{2}{c}{Protected} & \multicolumn{2}{c}{DTLS} & \multicolumn{2}{c}{OSCORE} & \multicolumn{2}{c}{Protected} \\
    \cmidrule(lr){2-3}
    \cmidrule(lr){3-4}
    \cmidrule(lr){4-5}
    \cmidrule(lr){5-6}
    \cmidrule(lr){6-7}
    \cmidrule(lr){7-8}
    \cmidrule(lr){8-9}
    & Req & Resp & Req & Resp & Req & Resp & Req & Resp \\ \midrule
    Structure & -- & 0 & 11 & 11 & 4 & 3 & -- & 11\\
    Context ID & -- & 2 & 2 & 2 & 1 & 0 & -- & 1\\
    Nonce & -- & 2 & 8 & 8 & 1 & 0 & -- & 0\\
    MAC & -- & 8 & 8 & 8 & 8 & 8 & -- & 40\\
    \bottomrule
  \end{tabular}
\end{table}

For AES in CCM mode, the same nonce is required for encryption and decryption.
The nonce is of variable length and usually ranges between 7--13 bytes~\cite{RFC-3610}.
We design our experiment to use partially implicit nonces~\cite{RFC-5116}.
Two bytes of the nonce are encoded into messages, while the remaining bytes are deduced implicitly, \eg from the hash of a resource URI\@.
This allows $2^{16}$ messages per resource until a refresh of established security contexts is advisable.
OSCORE repeatedly encodes smaller values in a single byte and CoAP (Protected) uses a 2-byte representation.
In responses, OSCORE uses the same nonce to protect objects and thus omits the nonce.
CoAP over DTLS uses eight bytes as a result of concatenating the epoch and sequence number fields.
The remaining four bytes of the DTLS nonce are implicit and generated as part of the handshake process~\cite{RFC-6655}.
NDN benefits from the immutable content property: Since names always map to the respective content, its hash is used as nonce.

We use a message authentication code of eight bytes as defined by the TLS AES-CCM cipher suites~\cite{RFC-6655}.
NDN appends another 32-byte HMAC signature that envelops the complete response packet.

Table~\ref{tab:overheads} summarizes the message overhead for the discussed protocols.

\subsection{Security Overhead in Comparison to Basic CoAP and NDN~Messages}\label{sec:packet-structures}

We now dissect each message of the protocols under comparison in detail and relate the basic CoAP and NDN packet sizes to the security extension (see Figure~\ref{fig:packetstructure}).
Our analysis distinguishes between requests and responses and includes all handshake messages for DTLS\@.
We assume that a response payload includes a 2-byte temperature value.

IEEE~802.15.4 admits a maximum physical layer packet size of 127~octets.
Assuming a typical configuration of 8-byte source and destination hardware addresses, considering a given 2-byte frame control field, 1-byte sequence number, 2-byte PAN id, and a 2-byte frame check sequence, the total MAC header overhead adds up to 23 bytes for each protocol.
This leaves 104~bytes for upper layer headers and user data.

In CoAP setups, the 6LoWPAN header occupies 35 bytes because it accommodates three 6LoWPAN dispatch bytes and two IPv6 addresses.
Moreover, each packet counts six bytes for the compressed UDP header.

Special consideration is required for \textit{ClientHello} and \textit{ServerHello} packets in a DTLS handshake.
In contrast to previous calculation, they surpass the maximum physical packet size and trigger a hop-wise 6LoWPAN fragmentation.
While the MAC header overhead is therefore doubled, the 6LoWPAN overhead increases by only nine bytes for the inclusion of fragmentation dispatches in both fragments.

In contrast to unprotected CoAP responses, CoAP (Protected) messages inflate by 12 bytes to include the context id, nonce, and message authentication code of AES-CCM\@ (see Section~\ref{subsec:secoverhead}).
CoAP over DTLS emits 29 and 27 more bytes for requests and responses, respectively, due to the DTLS record layer.
OSCORE messages display similar but extenuating effects: requests increase by 14 and responses by only 11 bytes.
The primary explanation for this surprisingly smaller increase is a reduced header overhead of OSCORE compared to the DTLS~1.2 record layer.
Nonces are further omitted from responses to decrease their header overhead.

In contrast to CoAP, where responses display smaller packet sizes than requests, NDN data packets exhibit larger sizes than Interests.
This is a result of names being fully included in returning data packets.
NDN data packets increase by an 8-byte AES-CCM MAC and an 32-byte HMAC signature, compared to unsecured NDN packets with an overall packet size of 64~bytes.
Since Interest messages do not contain any security measures, their packet sizes remain unaffected.


\section{Evaluation in the Testbed}\label{sec:eval}

In this section, we compare the different protocol configurations based on real implementations deployed in a testbed.

\subsection{Experiment Setup}
\paragraph{Scenarios \& Parameters}
We want to quantify the performances of a protected CoAP and NDN communication in a typical IoT data collection scenario with multiple sensor nodes.
For this, subsequent requests periodically traverse a gateway into an IoT stub network.
Each sensor device is requested 1000 times at an interval of $2\pm 0.5s$ and returns a 2-byte temperature reading.
To allow for comparison of pull-based NDN with CoAP, we limit CoAP methods to confirmable GET\@.

We align our experiments with respect to retransmission and timeout configurations.
All protocols employ the same retransmission strategy: On failures, nodes wait two seconds before retransmitting the original request.
In NDN, retransmissions are performed hop-by-hop, while CoAP performs them end-to-end.
At most four retransmissions will occur for each data.

We do not consider congestion from external cross-traffic in this work.
However, each individual transmission experiences self-induced background traffic from on-going requests and retransmissions.
The jittered request interval further mixes the event space and allows a greater exploration of the state space.
On average, this cross-traffic is constant per experimental run.

\paragraph{Software \& Hardware Platform}
All devices run RIOT version 2019.10.
NDN deployments are based on CCN-lite, and CoAP experiments use the default GNRC network stack of RIOT including libOSCORE and tinyDTLS (\cf Section~\ref{sec:implementation}).

\begin{figure}[]
  \centering
  \includegraphics{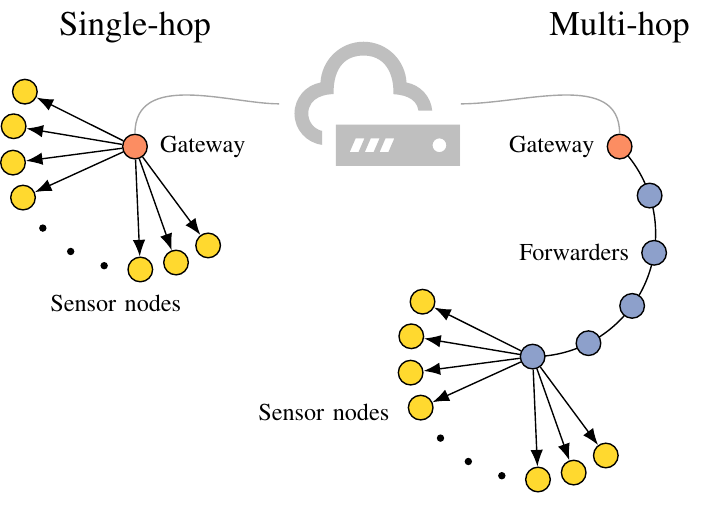}%
  \caption{Topologies for single-hop and multi-hop experiments.}%
  \label{fig:exp-setup}
\end{figure}

We conduct all experiments on the FIT IoT-LAB~\cite{abfhm-filso-15} testbed.
The hardware platform consists of class~2 devices~\cite{RFC-7228} featuring an ARM Cortex-M3 MCU with 64~kB of RAM and 512~kB of ROM\@.
Each device is equipped with an Atmel AT86RF231~\cite{a-lptzi-09} transceiver to operate on the IEEE 802.15.4 radio.
The testbed provides access to several sites with varying properties.
We perform our experiments on the \textit{grenoble} site in a single-hop and multi-hop configuration.
Our single-hop setup consists of one gateway node and ten sensor nodes in broadcast range as illustrated in Figure~\ref{fig:exp-setup}.
In the multi-hop configuration, we use one gateway node, ten sensor nodes, and five forwarder nodes.
Forwarding states are statically configured on each node to form the topology depicted in Figure~\ref{fig:exp-setup}.

\paragraph{Protocol Configurations \& Start-Up Conditions}
In all setups, we use AES in CCM mode with a 128-bit key and limit the resulting message authentication code to eight bytes as described in~\cite{RFC-6655}.
Each configuration also contains a 1-byte key identifier where applicable.
The NDN (Protected) setup further includes a hash-based message authentication code (HMAC) salted with a pre-shared key.
We limit the number of security contexts on the gateway to ten and on each sensor node to one.
As a consequence, sensor devices maintain only one DTLS session concurrently and all secured content objects from a particular sensor device use a single security context.
We configure all compared security protocols with pre-shared keys and context related variables, such as sequence counters, are set to default on device start-up.

\subsection{Time to Completion}
\begin{figure}
  \centering
  \includegraphics{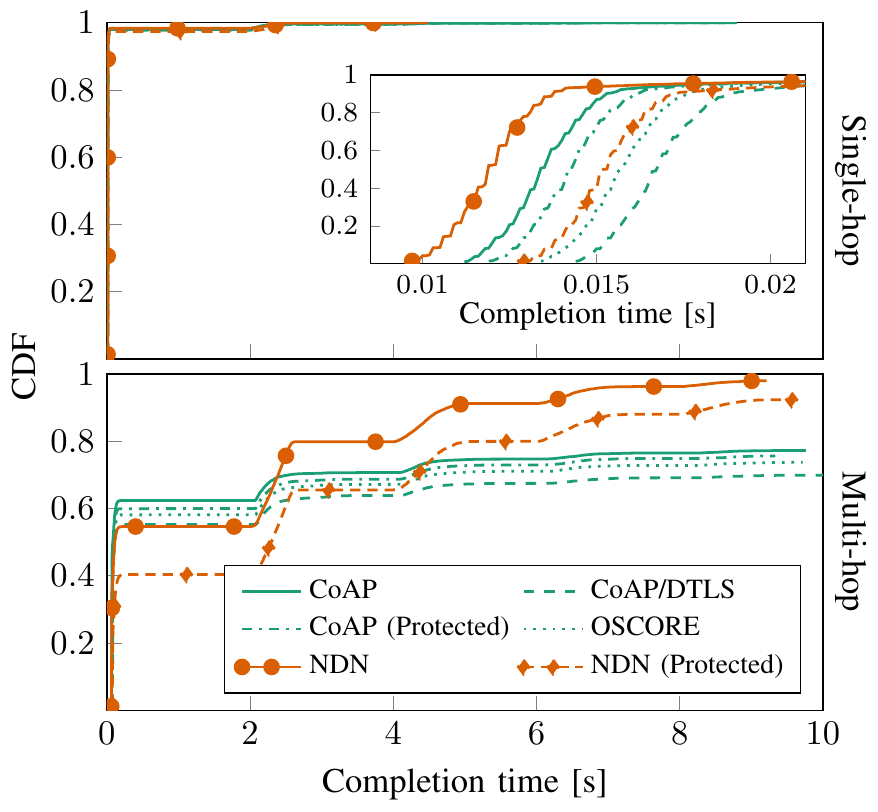}%
  \caption{Temporal distributions of content arrival times.}%
  \label{fig:reqresptime-cdf}
\end{figure}
\begin{figure*}
  \centering
  \includegraphics{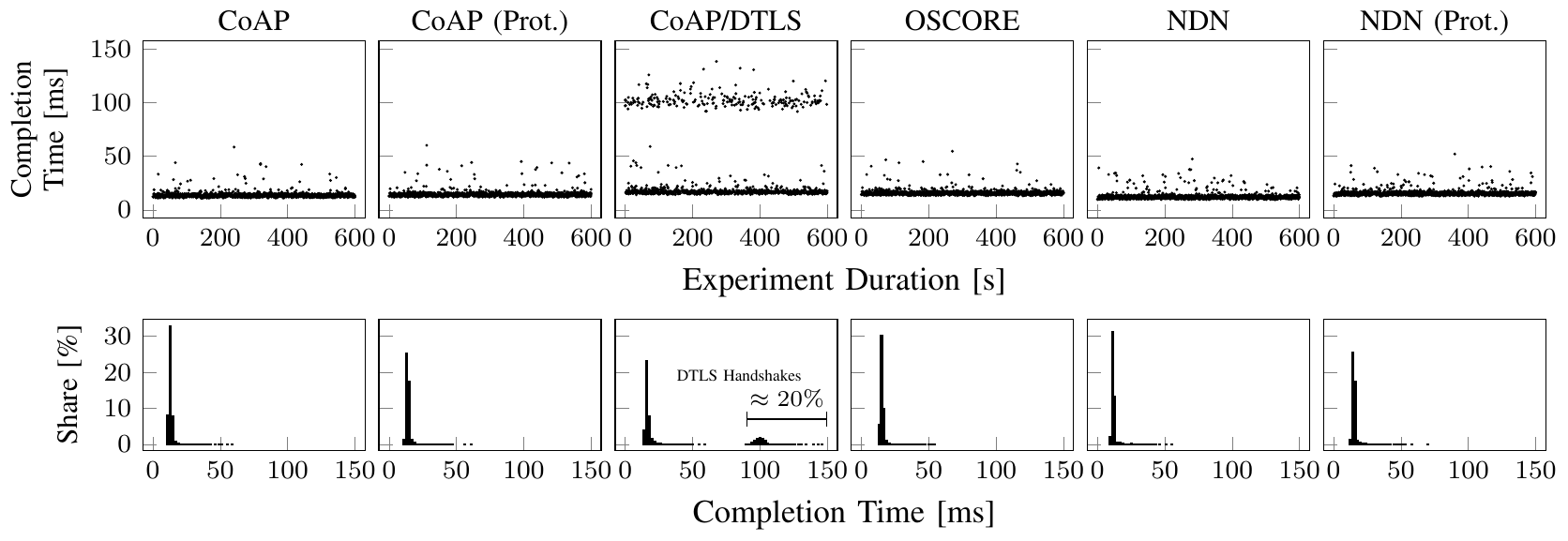}%
  \caption{Content arrival times and their percental distribution during the evolution of an experiment in a single-hop scenario.}%
  \label{fig:reqresptime}
\end{figure*}

We examine the delays measured between content requests and content arrivals at the gateway.
Figure~\ref{fig:reqresptime-cdf} combines the results for CoAP and NDN configurations in the single-hop and multi-hop setup.
We first observe that protocol families are in rough accordance in the single-hop case.
Temporal performances indicate a subsecond completion time for close to 100\% of all transmissions across the protocols.
The unprotected NDN configuration displays the fastest operation with 50\% of transmissions finishing below 11~ms.
Combining this observation with our previous result that NDN transmits the smallest request and response messages (see Figure~\ref{fig:packetstructure}), we can assume that unprotected NDN operates quickly.
In the unprotected CoAP configuration, 50\% of transmissions finish below 13~ms.
The protected protocol versions follow closely, whereby CoAP over DTLS is on the slow end with 50\% of transmissions finishing below 16~ms.

We then consider the more challenging multi-hop scenario.
Overall, results reveal a much slower protocol operation.
This reflects the common experience in low-power regimes that radio interferences and individual error probabilities accumulate over several hops and decrease global reliability.
The staircase pattern visible for each protocol is based on request retransmissions and the configured interval of 2~s per retransmission.
On the slower end, stairs show attenuating effects due to an accumulating jitter for each retransmission.
We observe that all CoAP variants operate in agreement.
Roughly 55--60\% of content requests complete in the subsecond range without requiring retransmissions.
Corrective actions delay the completion time, but are able to increase the number of successful responses at the gateway to 70--77\%.
The effects of inflated messages also become apparent:
CoAP emits the smallest packets and reveals a better performance than CoAP over DTLS, which emits the largest packets.
The delay distributions for NDN show surprising results.
Approximately 40--55\% of all responses arrive in the subsecond range at the gateway and therefore may indicate worse performances for NDN compared to the CoAP variants.
The determining factor for this discrepancy is due to the different retransmissions strategies.
NDN implements hop-wise retransmission, which raises the number of packets on a forwarding path.
This leads to more likely interferences that affect ongoing and subsequent content requests.
The effect of hop-wise retransmissions, however, is two-sided and unfolds advantageous effects when combined with in-network caching.
Corrective actions are able to boost the overall reliability of the NDN family to 92--97\%.

\subsection{Security Setup Overhead}

We now inspect the case where an endpoint repeatedly connects to the IoT stub network to retrieve sensor readings.
This setup follows the previous single-hop scenario with one minor change:
The endpoint at the gateway executes the steps necessary for entering a deep sleep mode after 20\% of all exchanges.
Figure~\ref{fig:reqresptime} summarizes the evolution of content arrival times throughout an experiment duration of ten minutes.
In the top row, we observe time to completion measurements for each protocol, while the bottom row visualizes the percental distribution for measured completion times.

Mostly the completion times reflect the same results as presented in Figure~\ref{fig:reqresptime-cdf}.
An obvious exception to this is CoAP over DTLS\@.
After a loss of library state at the gateway, a session handshake precedes the initial request.
A handshake for the configured cipher suite amounts to a total of ten DTLS packet transmissions.
Figure~\ref{fig:packetstructure} depicts the makeup of these packets and clearly shows that their sizes are comparable and in some cases even much larger than sizes of the actual authenticated and encrypted user packets.
Our evaluation shows that DTLS handshakes complete after around 100~ms, whereas in rare cases they even require up to 150~ms.
Since these handshakes take place between a single hop, their completion time serves as a good lower bound estimation for more complex scenarios.
Realistic multi-hop deployments with sufficient cross-side traffic and radio interferences are estimated to easily multiply these values for each added hop on a forwarding path.

A proper DTLS session resumption~\cite{RFC-5077} and Connection IDs~\cite{draft-ietf-tls-dtls-connection-id} could reduce the effects of handshakes after deep sleep or address changes, but are not available in tinyDTLS and require the persistence of context information that scales with the number of connected sensor devices.

\subsection{Request Creation Time}
\begin{figure}
  \centering
  \includegraphics{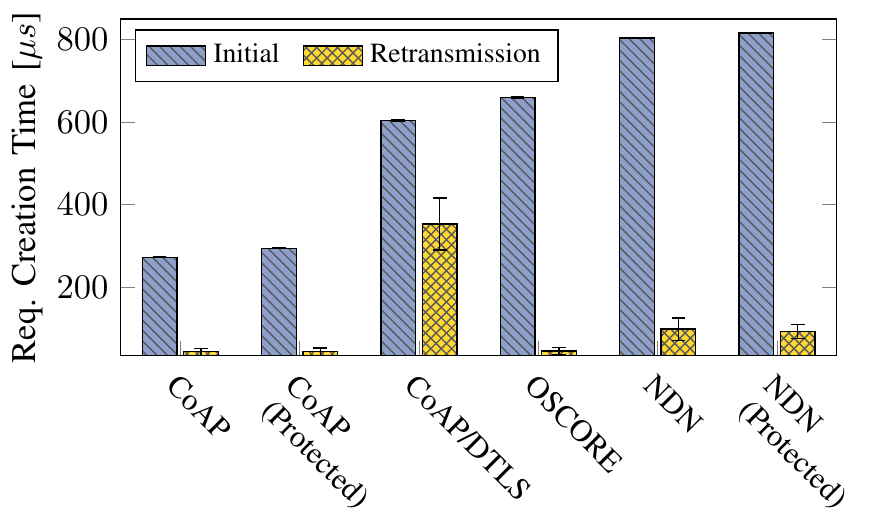}%
  \caption{Creation time for initial and retransmitted requests.}%
  \label{fig:reqresptime-retrans-bar}
\end{figure}

Our next evaluation gauges the message creation time of requests.
We measure the time starting from when an application triggers a request and stop it as soon as the packet is about to be passed to the next layer, which is UDP in the CoAP variants and the link-layer in NDN\@.
Figure~\ref{fig:reqresptime-retrans-bar} visualizes the results using two bar plots for each protocol.
The first bar shows average creation times for initial requests and the second bar denotes average creation times for request retransmissions.

CoAP in its unprotected and protected configurations exhibits the lowest creation times at around 280~$\mu$s.
Since the protected version only affects responses, equal times for requests are expected.
In both protocol versions, retransmissions are built much quicker in around 43~$\mu$s, since they already exist in retransmission buffers.
NDN behaves similarly, but creation times increase to $\approx$ 810~$\mu$s for initial requests and $\approx$ 95~$\mu$s for request retransmissions.
CoAP over DTLS behaves differently but mostly as expected.
Initial request creation times escalate to around 600~$\mu$s due to the authenticated encryption.
Interestingly, request retransmissions reveal unexpected results.
Requests exist in CoAP retransmission buffers and reduce overall creation times, but they still require to pass through the DTLS layer.
Hence, retransmission creation times spend on average around 353~$\mu$s and therefore eight times the amount of other CoAP setups.
This problem arises because of layering different protocols, which is not present in the OSCORE\@.
Protection takes place on the CoAP layer and retransmission buffers already contain protected messages.
Creation times for retransmissions reside at around 45~$\mu$s and are thus comparable to the protected and unprotected CoAP composition.


\section{Conclusions and Outlook}\label{sec:conclusion}
In this paper, we presented the first comprehensive analysis of protocols that aim for securing content  in the Internet of Things, with a special focus on constrained, low-power networks.
We conducted a theoretical and an experimental study, which compared different CoAP configurations, OSCORE, and named-data networking.
Selecting these protocols, we spanned the full solution space from end-to-end security sessions that act on transport channels to approaches that secure each data chunk individually.

Our findings indicate that in simple single-hop scenarios both security paradigms perform similarly in terms of content delivery.
Surprisingly, OSCORE, which is integrated into the CoAP mechanics, exhibits a much smaller security header overhead. This results in smaller packets and thus reduces latencies compared to the common CoAP over DTLS setup.
In challenged multi-hop topologies, we observed significant distinctions between the different protocols.
NDN benefits from both hop-wise transfer and in-network caches, which increase the reliability of data delivery remarkably and thus outperforms the other approaches.

As we show both the impact of overheads and of hop-by-hop retransmissions,
these results help justify, guide and (in future work) evaluate further improvements in the compared protocols:
The upcoming DTLS~1.3 will optimize the record layer footprint~\cite{draft-mattsson-lwig-security-protocol-comparison}.
OSCORE extensions that allow comparing the effects of unanticipated shutdowns (OSCORE replay window recovery~\cite[Appendix B.1]{RFC-8613} or LAKE~\cite{draft-ietf-lake-reqs}) will need to keep extra round-trips to a minimum.
Cacheable group observations~\cite{draft-tiloca-core-observe-multicast-notifications} could bring NDN-like features to CoAP, which would be beneficial as discussed in this~paper. We will analyze these emerging approaches in future work.


\paragraph{A Note on Reproducibility}
We fully support reproducible research~\cite{acmrep,swgsc-terrc-17} and perform all our experiments using open source software and an open access testbed.
Code and documentation will be available on Github at \url{https://github.com/inetrg/IFIP-Networking-2020}.

\paragraph{Acknowledgements}
This work was supported in part by the German Federal Ministry for Education and Research (BMBF) within the projects {\em I3 -- Information Centric Networking for the Industrial Internet}, {\em RAPstore -- RIOT App Store}, and the Hamburg {\em ahoi.digital} initiative with {\em SANE}.
The \texttt{libOSCORE} library was made with financial support from Ericsson AB.


\balance

\bibliographystyle{IEEEtran}
\bibliography{own,rfcs,ids,theory,complexity,layer2,internet,transport,overlay,hypermedia,vcoip,visualization,security,ngi,iot,meta}

\end{document}